\documentclass[useAMS,usenatbib,aas_macros]{mn2e}
\usepackage{aas_macros}
\usepackage{amsmath}

\usepackage{color}

\newcommand{\equ}[1]{eq.~(\ref{eq:#1})}
\newcommand{\equs}[1]{eqs.~(\ref{eq:#1})}

\newcommand{\Equ}[1]{Eq.~(\ref{eq:#1})}

\newcommand{\se}[1]{\S\ref{sec:#1}}
\newcommand{\fig}[1]{Fig.~\ref{fig:#1}}

\newcommand{\Fig}[1]{Figure~\ref{fig:#1}}

\newcommand{\tab}[1]{Table~\ref{tab:#1}}
\newcommand{\be}{\begin{equation}}
\newcommand{\ee}{\end{equation}}
\newcommand{\bea}{\begin{eqnarray}}
\newcommand{\eea}{\end{eqnarray}}

\def\la{\langle}

\newcommand{\msun}{M_\odot}

\newcommand{\ifm}[1]{\relax\ifmmode#1\else$\mathsurround=0pt #1$\fi}
\newcommand{\kms}{\ifmmode\,{\rm km}\,{\rm s}^{-1}\else km$\,$s$^{-1}$\fi}

\newcommand{\Gyr}{\,{\rm Gyr}}

\newcommand{\ltsima}{$\; \buildrel < \over \!\sim\! \;$}
\newcommand{\lsim}{\lower.5ex\hbox{\ltsima}}
\newcommand{\gtsima}{$\; \buildrel > \over \!\sim\! \;$}
\newcommand{\gsim}{\lower.5ex\hbox{\gtsima}}

\def\Mv{M_{\rm v}}

\def\Mg{M_{\rm g}}
\def\Ms{M_{\rm s}}
\def\Ma{M_{\rm a}}

\def\Mb{M_{\rm b}}

\def\Rd{R_{\rm d}}

\def\Vd{V_{\rm d}}
\def\fg{f_{\rm g}}

\def\Vd{V_{\rm d}}

\def\Mloss{M_{\rm loss}}
\def\Mdotloss{\dot{M}_{\rm loss}}

\def\Mdotout{\dot{M}_{\rm out}}

\def\Mdotrec{\dot{M}_{\rm rec}}
\def\Mdotg{{\dot{M}_{\rm g}}}
\def\Mdots{{\dot{M}_{\rm s}}}
\def\Mdota{{\dot{M}_{\rm a}}}
\def\Mdotv{{\dot{M}_{\rm v}}}
\def\etaout{{\eta_{\rm out}}}
\def\etarec{{\eta_{\rm rec}}}

\def\td{t_{\rm d}}

\def\tsf{t_{\rm sf}}

\def\fb{f_{\rm b}}

\def\Mdotg{\dot{M}_{\rm g}}
\def\Mdota{\dot{M}_{\rm a}}
\def\Mdotsf{\dot{M}_{\rm sf}}
\def\Mdots{\dot{M}_{\rm s}}

\def\fga{f_{\rm ga}}
\def\fsa{f_{\rm sa}}
\def\fg{f_{\rm g}}
\def\fs{f_{\rm s}}
\def\fgt{\varphi_{\rm g}}
\def\fst{\varphi_{\rm s}}
\def\fbt{\varphi_{\rm b}}
\def\stv{f_{\rm sv}}

\def\ta{t_{\rm a}}
\def\td{t_{\rm d}}
\def\tsf{t_{\rm sf}}
\def\ti{t_{\rm i}}
\def\zi{z_{\rm i}}

\def\Ma{M_{\rm a}}
\def\Mai{M_{\rm ai}}

\def\p{p}

\title[Minimal bathtub model]{An analytic solution for the minimal bathtub 
toy model: challenges in the star-formation history of high-z galaxies}
\author[A. Dekel and N. Mandelker]
{
Avishai Dekel$^{1}$, Nir Mandelker$^{1}$ 
\\
\\
$^1$Center for Astrophysics and Planetary Science, Racah Institute of Physics, 
The Hebrew University, Jerusalem 91904 Israel;\\
(dekel@huji.ac.il)
}
\begin{document}

\large
%\Large
%\LARGE

\pagerange{\pageref{firstpage}--\pageref{lastpage}} \pubyear{2002}

\maketitle

\label{firstpage}

\begin{abstract}
We study the minimal ``bathtub" toy model as an analytic tool for capturing key
processes of galaxy evolution and identifying robust successes and challenges 
in reproducing observations at high redshift.  The source and sink terms of the
continuity equations for gas and stars are expressed in simple terms from first
principles.  The assumed dependence of star-formation rate (SFR) on gas mass 
self-regulates the system into a unique asymptotic behavior, which is 
approximated by an analytic quasi-steady-state solution (QSS).  We address the 
validity of the QSS at different epochs independent of earlier conditions.  At 
high $z$, where the accretion is assumed to consist of gas only, the specific 
SFR is robustly predicted to be sSFR$\,\simeq[(1+z)/3]^{5/2}\Gyr^{-1}$, 
slightly higher than the cosmological specific accretion rate, in agreement 
with observations at $z=3\!-\!8$.  The gas fraction is expected to decline 
slowly, and the observations constrain the SFR efficiency per dynamical time to
$\epsilon\!\simeq\!0.02$. The stellar-to-virial mass ratio $\stv$ is predicted 
to be constant in time, and the observed value requires an outflow mass-lading 
factor of $\eta\!\simeq\!1\!-\!3$, depending on the penetration efficiency of 
gas into the galaxy.  However, at $z\!\sim\!2$, where stars are also accreted 
through mergers, the simplest model has an apparent difficulty in matching 
observations.  The model that maximizes the sSFR, with the outflows fully 
recycled, falls short by a factor $\sim\!3$ in sSFR, and overestimates $\stv$. 
With strong outflows, the model can reproduce the observed $\stv$ but at the 
expense of underestimating the sSFR by an order of magnitude.  We discuss 
potential remedies including a bias due to the exclusion of quenched galaxies.
\end{abstract}


%%%%%%%%%%%%%%%%%%%% 1
\section{Introduction}
\label{sec:intro}

% bathtub in words
The bathtub toy model is simply the equation of conservation of gas 
mass in a galaxy, or in one of its components, sometimes combined with the 
analogous equation for stellar mass. The gas equation describes the net 
change of gas mass in the inter-stellar medium (ISM)
as a difference between source terms and sink terms. 
When a whole galaxy is considered,
the source term is the accretion rate, dictated by cosmology,
but it can include recycled gas returning to the galaxy. 
The main sink terms are the rate at which gas turns into stellar mass
and the gas outflow rate.
The key for making the solution of this equation converge to a unique solution
independent of the initial conditions is that the SFR is assumed to be
proportional to the gas mass, and the other sink terms, especially the outflow
rate, are assumed to be proportional to the SFR. This generates an interplay
between the two components, where more gas mass allows a higher SFR which in
turn reduces the gas mass, thus driving the system into a self-regulated state
that is determined by the relative efficiencies of accretion, SFR 
and outflows.
When the accretion rate and the SFR timescale vary slowly enough, 
the asymptotic solution can be approximated by a quasi-steady state (QSS,
sometimes termed ``equilibrium") that can be derived analytically.

% attraction
\smallskip 
The bathtub model is useful because it captures the key processes
in a very simple and transparent way that makes it easy to trace their roles
in the global evolution. 
It is appealing because it converges to a unique attractor solution,
and even more so because of the analytic QSS solution.
One of its attractive features is that at any time 
this solution is unique, independent of the initial conditions and
independent of whether the same model was valid at earlier epochs.
The model is thus useful even if its ingredients with specific values
of the model parameters are valid only in a relatively short cosmological 
time interval.

% earlier usage
\smallskip
The bathtub model has become a very useful tool 
in understanding in simple terms the gross features of galaxy evolution.
As summarized in \citet{dekel13},
it has been applied in different ways to study different aspects of the
the evolution of a whole galaxy
\citep{bouche10,dave11I,dave12,kd12,lilly13,birrer_lilly14}.
Alternatively, it has been applied to the evolution of discs undergoing violent
disk instability (VDI) where another sink term is the mass inflow from the 
disc to the central bulge \citep{dsc09,cdg12,gdc12,forbes13}.
Recently it has been applied to the mass evolution of individual
giant clumps in VDI disks \citep{dekel_mass_14}.

% toy model vs SAM and sims
\smallskip
The bathtub toy model is not a replacement for the more elaborate
semi-analytic models (SAMs) or for full hydro-cosmological simulations.
This is a toy model, where on one hand the physical processes are represented
by very idealized recipes but on the other hand their effects are fully 
transparent. 
It can thus serve for solid qualitative constraints on the parameters that
characterize the accretion, SFR and outflows.
It also serves for revealing robust successes or difficulties in
understanding the origin of observed galaxy properties.

% in this paper study model
\smallskip
In the first part of this paper,
we present the bathtub toy model and its ingredients in the case of
a galaxy accreting from the cosmic web. 
We express the source and sink terms using the simplest possible scaling
relations that we try to motivate from first principles rather than from 
simulations or observations. These simplified approximate expressions,
in the Einstein-deSitter cosmological regime (approximately valid at $z>1$),
allow an analytic derivation of the QSS solution.
We numerically compute the asymptotic solution, 
derive the analytic QSS approximation, 
analyze its range of validity, and reveal the
role played by each process in major observable quantities at different
cosmological epochs.

% obs
\smallskip
In the second part we address three rather intriguing observational results
at high redshift, which we describe in more detail in \se{comp_obs}. 
First is the average specific star-formation rate (sSFR)
for massive star-forming galaxies (SFGs) as a function of redshift,
which current estimates show to decline from 
$\sim\! 10 \Gyr^{-1}$ at $z \!\sim\! 8$ through 
$\sim\! 2 \Gyr^{-1}$ at $z \!\sim\! 2$ to $\sim\! 0.1-0.2 \Gyr^{-1}$ at $z=0$
\citep{whitaker12_ssfr,stark13,labbe13,salmon14}.
Second is the gas fraction in the galaxy, $\fg$, deduced to be $\sim\! 0.5$ at 
$z \!\sim\! 2$ and rather slowly declining with time at high redshifts
\citep{daddi10,tacconi13}. 
Third is the stellar-to-virial mass ratio, $\stv$, estimated to be 
independent of redshift at $z=0-4$, with a peak of $\sim\! 0.025$ 
for halo masses in the range $(0.5-5)\times 10^{12}\msun$ 
\citep[e.g.][see \se{comp_obs}]{moster10,moster13,behroozi13}.
Given the universal baryon fraction $\fb \simeq 0.17$, this implies that 
only $\sim\! 0.15$ of the baryons that were supposed to enter the halo have 
made it to the central galaxy.
The sSFR that exceeds the expected specific accretion rate at all times,
the high gas fraction as late as $z \!\sim\! 2$, and the very low 
$\stv$, all pose non-trivial theoretical challenges.

\smallskip
In addition, the observed Kennicutt-Schmidt star-formation law
\citep{kennicutt89},
which can be interpreted as a relation between the SFR in a galaxy and its gas
content, reveals a rather constant SFR efficiency in a free-fall time
at the level of one to a few percent \citep{krum_tan07,krum_thomp07,
daddi10_sfr,genzel10,kdm12}.
Then,
significant outflows are observed in high-redshift galaxies
\citep{steidel10,genzel11}, with mass-loading factors ranging from below unity
to 10. 
In both cases the uncertainties are large, so we prefer to 
treat these as free variables.

\smallskip
The paper is organized as follows.
In \se{model} we present the bathtub model and its ingredients. 
In \se{qss} we describe the quasi-steady-state solution and work out 
analytically observable quantities of interest.
In \se{validity} we study the validity of the QSS solution.
In \se{comp_obs} we make a simple comparison of the model to observations.
In \se{disc} we discuss the successes and failures of the model
in reproducing the observations,
and address the assumptions made in the toy model.
In \se{conc} we conclude our findings concerning the toy model and the 
comparison to observations.

%%%%%%%%%%%%%%%%%%%%%
\section{The Minimal Bathtub Model}
\label{sec:model}

%-----------------
\subsection{Continuity Equations}
\label{sec:cont}

The bathtub model consists of
straightforward mass conservation for the gas mass $\Mg$
and the stellar mass $\Ms$ in an evolving galaxy,
\be
\Mdotg = \fga \Mdota - (\mu+\eta)\, \Mdotsf \, ,
\label{eq:gas}
\ee 
\be
\Mdots = (1-\fga)\, \Mdota + \mu\, \Mdotsf \, .
\label{eq:stars}
\ee
The accretion-rate term $\Mdota$ refers to {\it all} the baryons, gas and 
stars, as they {\it enter the galaxy for the first time}, where
$\fga$ and $\fsa=1-\fga$ are the fractions of gas and stars in this
accretion, respectively (to be spelled out in \se{ar}). 
The term $\Mdotsf$ is the star-formation rate (SFR) (\se{sfr}).
The parameter $\mu$ is the fraction of mass in forming stars that remains 
in stars, the rest assumed to be instantaneously lost from the stars due 
to supernovae and stellar winds and deposited back in $\Mg$ \citep{tinsley80},
see \se{sfr}.
The parameter $\eta$ is the {\it effective} mass-loading factor of the 
gas loss from the galaxy, 
\be
\eta = \frac{\Mdotloss}{\Mdotsf} \, ,
\label{eq:eta}
\ee
referring to the net gas lost from the galaxy $\Mloss=\Ma-\Mb$,
where $\Ma$ is the total fresh baryonic mass accreted into the galaxy,
and $\Mb=\Mg+\Ms$ is the baryonic mass in the galaxy.
In our definition of the effective $\eta$,
the mass loss is the outflow driven by stellar or AGN feedback
minus the gas that flew out earlier and is now returning into the galaxy,
termed ``recycling" (\se{outflow}).

%------------------
\subsection{Accretion Rate}
\label{sec:ar}

The average specific accretion rate (sAR) of baryons into the galaxy, 
$\Mdota/\Ma$,
is approximated by the specific accretion rate of total matter into the 
virial radius of the dark-matter halo, $\Mdotv/\Mv$,
where $\Mv$ is the total virial mass. 
The latter has been estimated quite robustly using theoretical considerations,
confirmed and fine-tuned by simulations \citep{neistein08}.
The estimate, based on very simple arguments in the  EdS regime,
gives \citep{dekel13}
\be
\frac{\Mdota}{\Ma} \simeq s\, (1+z)^{5/2}  \Gyr^{-1} \, ,
\quad s \simeq 0.03 \Gyr^{-1} \, .
\label{eq:Mdota}
\ee
The power $5/2$ is exact in the EdS regime, 
stemming from the fact that $w \propto (1+z) \propto t^{-2/3}$ 
is a self-invariant time variable for structure formation,
namely the halo-mass growth rate $d\Mv/dw$ is constant in time, so 
$\dot \Mv \propto \dot w$ with $\dot w \propto t^{-5/3} \propto (1+z)^{5/2}$.
The slightly smaller powers proposed elsewhere
\citep[e.g. 2.2-2.4 proposed in][]{neistein06,neistein08} 
meant to provide global best fits in a larger redshift range, 
including the acceleration phase at $z<1$.
Similar fitting formulae based on simulations were proposed by others
\citep{fakhouri08,genel08,genel10}.

\smallskip
The approximation in \equ{Mdota} ignores a weak additional mass dependence,
roughly proportional to $\Ma^{0.14}$ \citep{neistein08}.
The power $0.14$, which fits simulations for $\Mv$ in the range 
$10^{11}-10^{14}\msun$, originates from $(n+3)/6$ where $n\!\sim\!-2.1$ 
is the log slope of the linear fluctuation power spectrum on the relevant 
scales.
By ignoring this weak mass dependence one does not introduce a large error 
as long as the analysis involves a mass range limited to one or two decades. 
This approximation is adopted in our minimal toy model for two important 
reasons, as follows.

\smallskip
First, having the specific accretion rate 
independent of mass makes the sAR adopted in \equ{Mdota} the same at the halo
virial radius and at the galaxy boundary in the inner halo.
We assume that the fraction of the baryons in the total mass accreted
(including dark matter) into the virial radius 
equals the universal baryonic fraction $\fb\!\simeq\!0.17$.
We denote by a {\it penetration} factor $\p$ the fraction of the baryons 
accreted into the virial radius
that actually penetrate through the halo into the central galaxy, 
\be
\p = \frac{\Ma}{\fb \Mv} \, .
\label{eq:p}
\ee
The sAR is independent of $\p$ because both $\Mdota$
and $\Ma$ are proportional to the same $\p$.
The above is confirmed in hydro-cosmological simulations
\citep[][Figures 5 and 10]{dekel13}, where $\p$ is found to be about 0.5.
Thus, the minimal toy model with \equ{Mdota} is independent of 
the actual value of $\p$.
However, $\p$ will enter when we compare the model predictions to the
observed stellar-to-virial ratio that involves $\Mv$.

\smallskip
The second benefit of the simple form of \equ{Mdota} is that it can be
integrated analytically to give a total baryon mass growth (ignoring outflows) 
of
\be
\Ma = \Mai\, e^{-\alpha(z-z_i)} \, ,
\quad \alpha = 1.5\,s\,t_1  = 0.79 \, ,
\label{eq:Ma}
\ee
where $\Mai$ is the initial value at time $\ti$ or redshift $\zi$.
Here we used the EdS approximation at $z>1$ \citep{dekel13},
\be
(1+z) = (t/t_1)^{-2/3} \, ,
\quad t_1 = 17.5 \Gyr \, .
\label{eq:t1}
\ee

\smallskip
At very high redshift, and for galaxies with relatively low masses,
one may assume in \equ{gas} $\fga=1$.
This has been the implicit assumption in several other applications of the
bathtub model \citep[e.g.][]{bouche10,dave12,lilly13}.
However, by $z\sim\! 2$ and for massive galaxies,
there is a non-negligible fraction of ex-situ stars coming in through
mergers, $\fsa >0$. 
It will turn out below, as has already been noticed by \citet{kd12}, 
that the value of $\fsa$ makes a significant difference.

% Dave 12, Lilly 13
\smallskip
It may be helpful to compare our current notation with that of
others who used the bathtub model, such as \citet[][D12]{dave12} and
\citet[][L13]{lilly13}.
For example, our $\mu$ is referred to in many cases as $1-R$.
Our penetration factor $\p$ is equivalent to $\zeta$ of D12 
(related to as ``the preventive feedback parameter"),
and to $f_{\rm gal}$ of L13.
In our case, however, $\p$ is not a basic parameter of the model. 
Our $\fst$ is the same as $f_{\rm star}$ in L12,
so our stellar-to-virial mass ratio $\stv=\p\fb\fst$ is in their notation 
$f_{\rm gal} \fb f_{\rm star}$.  %
The $\dot{M}_{\rm grav}$ of D12 is equivalent to our 
$\fga\fb\Mdotv=\fga\Mdota/\p$,
and their $\dot{M}_{\rm prev}$ would be expressed in our notation as
$(1-\p)\fga\fb\Mdotv$, where D12 and L13 implicitly assume $\fga\!=\!1$.  %
The quantity $\epsilon$ in L13 refers to the inverse of the depletion time,
SFR$/\Mg$, which in our notation is $\tsf^{-1}$,
while in our notation $\epsilon$ is the SFR efficiency in a dynamical time,
$\epsilon=\td/\tsf$, as more common in the literature.
Our fiducial case assumes that our $\epsilon$ is constant, while their
fiducial case is with their $\epsilon$ being constant.  %
Finally,
the modeling of recycling rate by D12 can be expressed as  
$\Mdotrec=[\alpha_Z/(1-\alpha_Z)]\,\Mdota$,
where $\alpha_Z$ is the ratio of metallicities in the inflowing and ambient 
ISM gas. This is actually equivalent to the way we model recycling,
except that they add it to the source term while we subtract it 
from the sink term, \se{outflow}.

\begin{table*}
\centering
  \begin{tabular}{lllc}
      \hline
  Quantity & Meaning & Definition & Fiducial value   \\
      \hline
 $\Mg,\Ms$  & gas, stellar mass in the galaxy   &           &     \\
 $\Ma$      & baryon mass accreted onto the galaxy &\equs{p} and (\ref{eq:Ma})
&
                                      \\
 $\Mdotsf$ & star-formation rate SFR              & \equ{sfr}   &  \\
      \hline
 $\eta$  & effective outflow mass-loading factor, $\eta=\etaout-\etarec$
                  & \equs{eta} and (\ref{eq:eta_out-rec}), \se{outflow} & 1 \\
 $\epsilon$  & SFR efficiency per dynamical time & \equ{sfr} & 0.02 \\
 $\mu$  & fraction of stellar mass formed that remained in stars $\mu=1-R$ &
                     \se{cont}, \se{sfr} & 0.54 \\
      \hline
 $s$    & average specific accretion rate at $z=0$ in $\Gyr^{-1}$ &
                 \equ{Mdota} & 0.03      \\
 $\fb$ & universal baryonic fraction           & \se{ar} & 0.17      \\
 $\fga$,$\fsa$ & gas, stellar fraction in the accretion
                            & \equs{gas}, (\ref{eq:stars})& 1,0 or 0.8,0.2 \\
 $\p$  & penetration factor                    & \equ{p} & 0.5       \\
      \hline
 $A$     & gas accretion rate $A=\fga \Mdota$ & \equ{equation} &           \\
 $\td$   & disc crossing time $\Rd/\Vd$       & \se{sfr} & \\
 $\nu$   & $\td$ in units of the cosmological time & \equ{td} & 0.0071  \\
 $\tsf$  & star-formation or depletion time   & \equ{sfr} &           \\
 $\tau$  & $\tau=\tsf/(\mu+\eta)$             & \equ{equation} &           \\
      \hline
 $\fg$,$\fs$  & gas, stellar fraction of baryons in the galaxy &\equ{fg_fs} &
\\
 $\fgt$,$\fst$  & gas, stellar fraction of baryons accreted $\Ma$ &
                           \equ{fgt_fst} & \\
 $\stv$  & stellar-to-virial mass ratio               & \equ{stv} &        \\
 sSFR  & specific star-formation rate                & \equ{sSFR} &        \\
      \hline
   \end{tabular}
  \caption{List of quantities and parameters}
\label{tab:parameters}
\end{table*}

%-------------------
\subsection{Star-Formation Rate}
\label{sec:sfr}

The key to a steady-state solution for \equ{gas}, motivated by the
empirical Kennicutt-Schmidt law and theoretical considerations 
\citep[e.g.][]{kdm12},
is that the SFR is assumed to be proportional to the gas mass,
\be
\Mdotsf = \frac{\Mg}{\tsf}\, , 
\quad \tsf = \epsilon^{-1} \td \, .
\label{eq:sfr}
\ee
The time $\tsf$ is the timescale for star formation to consume the gas
reservoir (ignoring $\mu$), 
also referred to as the {\it depletion} time \citep[e.g.][]{genzel08,dave12}. 
We assume here that it is proportional to the disk dynamical crossing 
time $\td =\Rd/\Vd$, where $\Rd$ and $\Vd$ are the characteristic radius
and rotation velocity of the disc \citep{kdm12}.
As in \citet{dekel13},
the disc dynamical time is approximated as proportional to the 
cosmological time,  
\be
\td = \nu\,t\, , \quad \nu \simeq 0.0071 \, .
\label{eq:td}
\ee
We assume that the SFR efficiency factor per dynamical time
$\epsilon$ is constant
over the time interval of interest and independent of galaxy mass 
in the mass range of interest.  
Its value is indicated to be on the order of $0.02$ \citep[e.g.][]{kdm12}.

\smallskip
% mu
A value of $\mu=0.54$ % 0.53
has been estimated in \citet[][Appendix A, where $R=1-\mu$]{kd12},
adopting a Chabrier IMF and assuming that stars in the ranges $(1-8)\msun$ 
and $>8\msun$ leave behind white dwarfs of $0.7\msun$ and neutron stars 
of $1.4\msun$ respectively. This estimate is valid after $z \sim 2$,
where the age of the Universe is several Gyr.
At higher redshifts, 
considering that only stars with lifetimes on the main sequence shorter than
3 Gyr (or 1 Gyr) had time to return mass to the ISM by $z=3$ (or $z=5.7$),
one obtains $\mu=0.57$ (or $0.62$).  
The maximum value relevant at very high redshifts is thus not very different
from the value of $0.54$ adopted throughout our current calculations.

%-----------------
\subsection{Outflow and Recycling}
\label{sec:outflow}

The outflow rate due to stellar feedback, $\Mdotout$, is assumed to be 
proportional to the instantaneous SFR, with a constant mass-loading factor 
$\etaout={\Mdotout}/{\Mdotsf}$.
Based on theory and observations of massive galaxies at $z\!\sim\! 2$,
$\etaout$ is expected to be of order unity or a few \citep{genzel11,dk13}.

\smallskip
A fraction of the gas that flew out at earlier times is assumed to be
returning at a recycling rate $\Mdotrec$. We crudely approximate this
also to be proportional to the instantaneous SFR, with a corresponding 
constant factor $\etarec={\Mdotrec}/{\Mdotsf}$.
The effective mass-loading factor is 
\be
\eta=\etaout-\etarec \, .
\label{eq:eta_out-rec}
\ee
When recycling is taken into account, 
the effective $\eta$ could be smaller than unity and even vanish.

\smallskip
In the quasi-steady-state solution discussed below (\se{qss}), the SFR becomes
proportional to the gas accretion rate, \equ{qss_Mdotsf}. 
In this case, the way we incorporate recycling
becomes equivalent to assuming that it is proportional to the instantaneous
accretion rate, and then adding the recycling to the source term instead of 
subtracting it from the sink term. The term to be added to the source term is 
$\Mdotrec=[\fga\etarec/(\mu+\eta)]\Mdota$.

%------------------------
\subsection{Summary: Parameters and Observables}
\label{sec:parameters}

A summary of the model parameters is as follows.
We adopt $s=0.03$, the value estimated for the average sAR.
We adopt $\mu \simeq 0.54$ as a rather robust estimate at $z<2$,
and recall that it could be up to $\sim 20\%$ higher at higher redshifts.
The stellar fraction in the accretion, $\fsa$, will turn out to be rather
important. It could be negligibly small at very high 
redshifts, say $z \geq 4$, but it is likely to be non-zero at intermediate
redshifts, $z \!\sim\! 2$.
Once the parameters mentioned above are fixed at their fiducial values,
the natural free parameters of the model are $\epsilon$ and $\eta$,
which could vary from below 0.01 to above 0.02, 
and from zero to more than a few, respectively.
In the current simplest version of the bathtub model, we assume that all these
parameters are mass independent in a certain relevant mass range, and
are constant during a certain cosmological time interval.

% penetration
The penetration parameter $\p$, which is necessary only for the comparison 
with observed fractions that involve the virial mass and is not an intrinsic
parameter of the model, could vary about a fiducial value of $\p \!\sim\! 0.5$.
This is based on hydro-cosmological simulations with a varying strength of
supernova and momentum-driven feedback 
\citep[][and work in preparation based on simulations with stronger 
feedback]{dekel13}.

\smallskip
For convenience, we define dimensionless fractions as follows.
The gas and stellar mass fractions with respect to the baryons
in the galaxy are denoted
\be
\fg = \frac{\Mg}{\Mb}\, , \quad
\fs = \frac{\Ms}{\Mb}\, , \quad
\fg+\fs=1 \, .
\label{eq:fg_fs}
\ee
The corresponding fractions with respect to all the baryons accreted
(including those that later flowed out) are
\be
\fgt = \frac{\Mg}{\Ma} , \quad
\fst = \frac{\Ms}{\Ma} , \quad
\fbt = \frac{\Mb}{\Ma} , \quad
\fgt+\fst=\fbt \, .
\label{eq:fgt_fst}
\ee

\smallskip
Most useful for characterizing the solution are the two
fractions $\fg$ and $\fst$,
both because each will turn out to be sensitive to a different parameter,
$\epsilon$ or $\eta$ respectively, 
and because they can be related to observations. 
The gas fraction $\fg$ is directly observable, and $\fst$ is related to the 
observable stellar-to-virial mass ratio
via the penetration parameter $\p$, \equ{p},
\be
\stv = \frac{\Ms}{\Mv} = \p\, \fb\, \fst \, .
\label{eq:stv}
\ee
Assuming initial values for $\fg$ and $\fst$ at an initial redshift $\zi$,
\equs{gas} and (\ref{eq:stars}) can be integrated forward in time.
The third and most useful quantity is the sSFR,
which is robustly predicted, insensitive to the values of $\epsilon$ and 
$\eta$, and is observable. 

% Notation Dave 12, Lilly 13
\smallskip
Most of the quantities and parameters used are listed for convenience in
\tab{parameters}.
It may be helpful to compare our current notation with that of
others who used the bathtub model, such as \citet[][D12]{dave12} and
\citet[][L13]{lilly13}.
For example, our $\mu$ is referred to in many cases as $1-R$.
Our penetration factor $\p$ is equivalent to $\zeta$ of D12
(related to as ``the preventive feedback parameter"),
and to $f_{\rm gal}$ of L13.
In our case, however, $\p$ is not a basic parameter of the model.
Our $\fst$ is the same as $f_{\rm star}$ in L12,
so our stellar-to-virial mass ratio $\stv=\p\fb\fst$ is in their notation
$f_{\rm gal} \fb f_{\rm star}$.  %
The $\dot{M}_{\rm grav}$ of D12 is equivalent to our
$\fga\fb\Mdotv=\fga\Mdota/\p$,
and their $\dot{M}_{\rm prev}$ would be expressed in our notation as
$(1-\p)\fga\fb\Mdotv$, where D12 and L13 implicitly assume $\fga\!=\!1$.  %
The quantity $\epsilon$ in L13 refers to the inverse of the depletion time,
SFR$/\Mg$, which in our notation is $\tsf^{-1}$,
while in our notation $\epsilon$ is the SFR efficiency in a dynamical time,
$\epsilon=\td/\tsf$, as more common in the literature. 
Our fiducial case assumes that our $\epsilon$ is constant, while their
fiducial case is with their $\epsilon$ being constant.  %
Finally,
the modeling of recycling rate by D12 can be expressed as
$\Mdotrec=[\alpha_Z/(1-\alpha_Z)]\,\Mdota$,
where $\alpha_Z$ is the ratio of metallicities in the inflowing and ambient
ISM gas. This is actually equivalent to the way we model recycling, 
except that they add it to the source term while we subtract it 
from the sink term, \se{outflow}.

% 1
\begin{figure*}
\vskip 7.1cm
\includegraphics{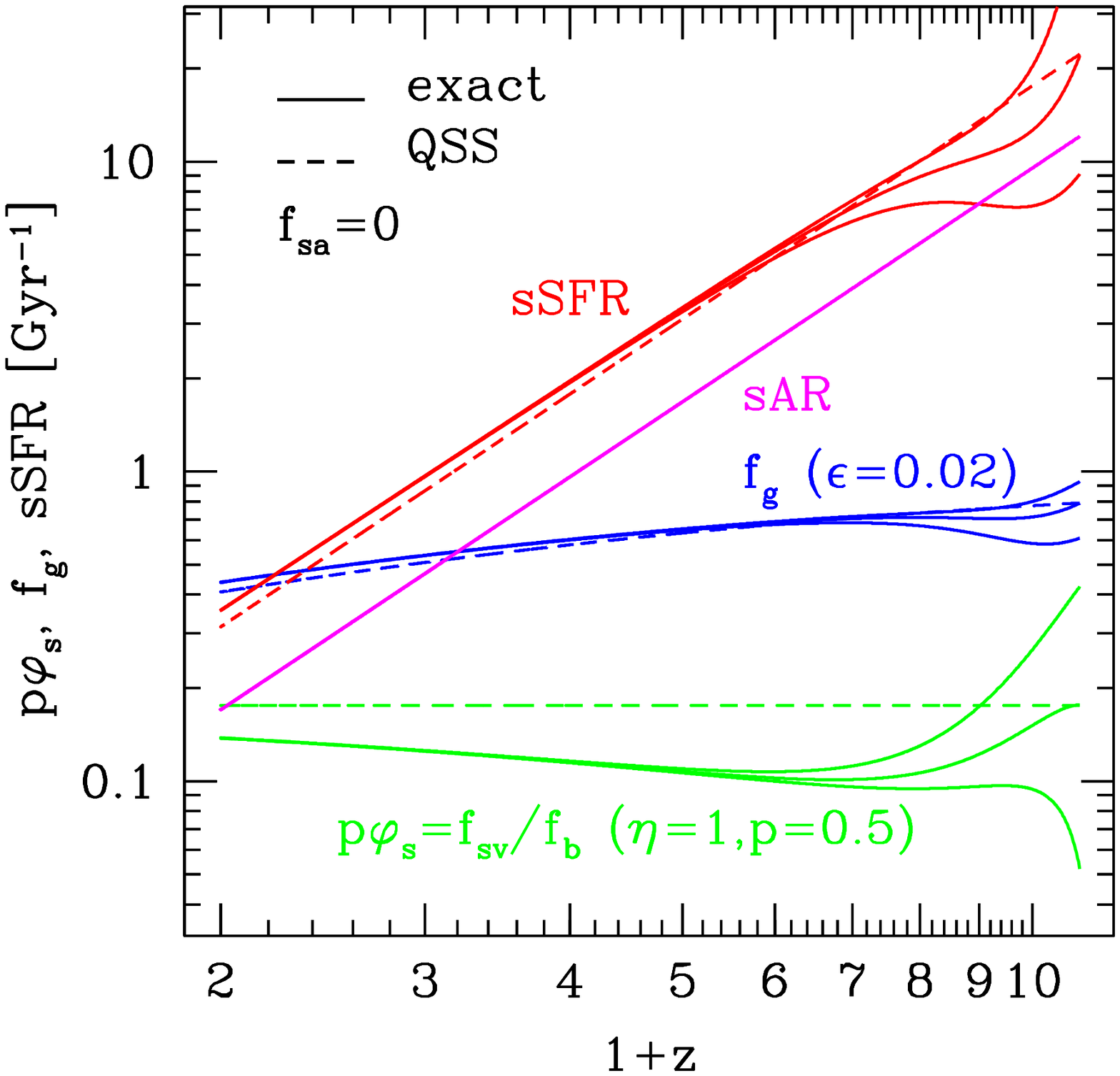}
\includegraphics{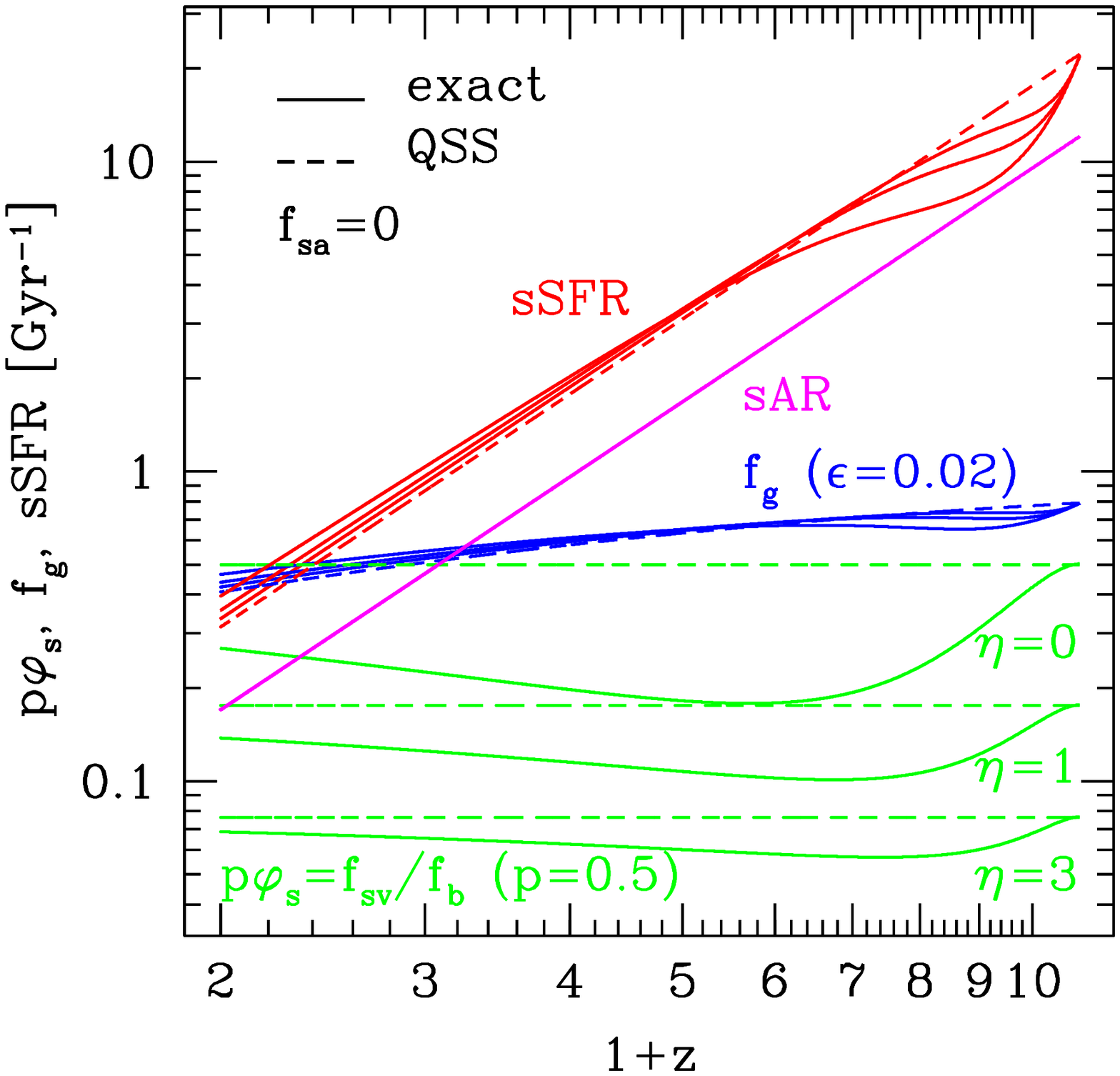}
\caption{
Solutions of the bathtub model as a function of $1+z$, representing
time evolution from $\zi=10$ to $z=1$.
Shown are the exact solution (solid) and the QSS toy-model prediction (dashed).
Shown are the sSFR (red) in comparison with the sAR,
the gas fraction in the galaxy $\fg$ (blue),
and $\p\fst$ with $\p=0.5$ (green), which is the stellar-to-halo mass ratio
divided by $\fb=0.17$, $\stv/\fb$.
The parameter choice here is $\fsa=0$ and $\epsilon=0.02$,
with $\mu=0.54$ and the average specific accretion rate (sAR, magenta)
with $s=0.03 \Gyr^{-1}$.
{\bf Left:}
For $\eta=1$, starting from three different initial conditions
($C/\Mai=0.5,0,-0.25$) at $\zi=10$.
The exact solution converges to a unique asymptotic solution independent of
the initial conditions on a timescale corresponding to $\Delta z \!\sim\! 3$.
{\bf Right:}
For initial conditions that obey the QSS solution, and $\eta=0,1,3$ (top
to bottom for the green curves, bottom up for the red and blue curves).
The QSS solution reproduces the exact solution for $\fg$ and sSFR very well.
Its deviation from the exact model for $\fst$ (green) is consistent with
\equ{rerror} -- it is decreasing with time and with $\eta$.
}
\label{fig:mgmst}
\end{figure*}

%%%%%%%%%%%%%%%%%%%%%%%%%%%%
\section{Quasi-Steady-State}
\label{sec:qss}

\smallskip
\Fig{mgmst} shows in solid lines the exact numerical solution of \equs{gas}
and (\ref{eq:stars}) for the
observable quantities, with the parameter choice $\fsa=0$ and $\epsilon=0.02$,
and with $\p=0.5$ multiplying $\fst$.
Shown for $\eta=1$ are cases with different initial conditions (left),
which converge to a unique asymptotic solution independent of the initial
conditions.
Also shown are solutions for different values of $\eta=0,1,3$ (right).
We next analyze the the convergence to the asymptotic solution and derive 
it analytically.

\subsection{Steady-State and Quasi-Steady-State}
\label{sec:qss_ss}

\Equ{gas} for $\Mg(t)$ is of the form 
\be
\Mdotg = A - \tau^{-1}\, \Mg \, ,
\label{eq:equation}
\ee
where the parameters are the gas accretion rate,
$A=\fga \Mdota$, and the characteristic timescale, $\tau=\tsf/(\mu+\eta)$.
The negative $\Mg$ term drives the system into a unique attractor solution.

\smallskip
If $A$ and $\tau$ are both constant in time, there is a simple solution to 
this equation, 
\be
\Mg(t)= A\,\tau\,(1-e^{-t/\tau})\, ,  \quad
\Mdotg = A\,e^{-t/\tau} \, ,
\label{eq:ts}
\ee
for the initial condition $\Mg=0$ at $t=0$.
For $t \gg \tau$, the transient component decays exponentially and the solution 
converges asymptotically to the steady-state (SS) solution
\be
\Mg=A\,\tau \, , \quad  \Mdotg=0 \, .
\label{eq:ss}
\ee
This is a stable attractive solution to which the solution 
converges independent of the initial value of $\Mg$ because $\Mdotg$ is a 
decreasing function of $\Mg$. 

\smallskip
In the cosmological case $A$ and $\tau$ are varying in time, but rather slowly.
If the variation timescale is significantly longer than $\tau$ (see
\se{validity}),
then one can approximate $A$ and $\tau$ as being constant during periods of 
order $\tau$ in which $\Mg$ is evolving rapidly,
and assume a temporary steady-state solution in the form of 
\equ{ss}.
%\footnote{This is analogous to the Born-Oppenheimer quantum analysis 
%of the hydrogen molecule, where the protons are assumed static when 
%calculating the electron dynamics.}
We thus approximate the asymptotic quasi-steady-state (QSS) solution at 
$t \gg \tau$ by
\be
\Mg(t) = A(t)\, \tau(t) = \frac{\fga\,\tsf(t)}{\mu+\eta}\,\Mdota(t)  \, . 
\label{eq:qss}
\ee
In the approximate QSS solution, \equ{gas} reads
\be
\Mg(t) = \frac{\theta(t)}{(\mu+\eta)}\,\Ma(t) \, ,
\label{eq:qss_Mg}
\ee
where
\be
\theta(t) = \fga\, \frac{\tsf(t)}{\ta(t)}\, , \quad
\ta = \frac{\Ma}{\Mdota} \, .
\label{eq:theta_def}
\ee
According to \equs{Mdota}, (\ref{eq:td}), and (\ref{eq:t1}),
$\theta$ evolves as
\bea
\theta(t)
&\!\!\! = \!\!\!& \fga\,s\,\nu\,t_1\,\epsilon^{-1} (1+z) \nonumber \\
&\!\!\! \simeq \!\!\!& 0.186\,\fga\, s_{0.03}\, \epsilon_{0.02}^{-1}\, (1+z)\,
.
\label{eq:theta}
\eea
According to the QSS solution, 
the SFR is proportional to the accretion rate,
\be
\Mdotsf=\frac{\fga}{\mu+\eta}\,\Mdota \, .
\label{eq:qss_Mdotsf}
\ee

%----------------
\subsection{QSS Solution for Observable Quantities}
\label{sec:qss_solution}

We next evaluate quantities of interest that also involve the stellar mass,
using the QSS solution of \equ{gas} combined with \equ{stars}.
We focus on the observable quantities, (a) the gas fraction $\fg$,
(b) the ratio of stellar mass to accreted baryon mass $\fst$ that is observable
through the stellar-to-virial mass given by $\stv=\p\fb\fst$, and
(c) the specific SFR,
\be
{\rm sSFR} = \frac{\Mdotsf}{\Ms} \, .
\label{eq:sSFR}
\ee

% solution for observables
Inserting the QSS solution of \equ{gas} into \equ{stars}, we obtain
\be
\Mdots = \frac{\mu+\fsa\,\eta}{\mu+\eta} \, \Mdota \, .
\label{eq:qss_Mdots}
\ee
If the parameters are all constant in time, and if the QSS solution is valid
since an initial time $\ti$, one can integrate this equation in time to obtain
\be
\Ms(t) = \frac{\mu+\fsa\,\eta}{\mu+\eta} \, \Ma(t)  +C \, ,
\label{eq:qss_Ms}
\ee
where $C$ is the constant of integration, determined by the initial conditions
at $\ti$.
%One can choose the ``initial" time $\ti$ such that $C=0$, 
%which is equivalent to
The choice $C=0$ is equivalent to
\be
\fbt(\ti) = \frac{\theta(\ti) + \mu+\fsa\eta}{\mu+\eta}\, .
\ee
Even if $C\neq 0$, since $\Ma$ grows rapidly according to \equ{Ma},
the role of $C$ in \equ{qss_Ms} becomes negligible a short while after $\zi$.
For example, $C/\Ma$ drops by a factor of 10 by
$z-\zi \simeq -2.3/\alpha \simeq -2.9$.
Thus, the effect of the initial conditions on the solution for $\Ms$
is expected to become negligible after a time interval corresponding to
$\Delta z \!\sim\! 3$, and the convergence is faster when the 
initial conditions do not deviate much from the asymptotic solution.

\smallskip
Adding the gas mass from \equ{qss} and the stellar mass from 
\equ{qss_Ms}, the baryonic mass is
\be
\Mb(t) = \frac{\theta(t)+\mu+\fsa\eta}{\mu+\eta}\, \Ma(t) \, .
\label{eq:qss_Mb}
\ee
Then, from \equs{qss} and (\ref{eq:qss_Mb}), the gas fraction is
\be
\fg = \frac{\theta(t)}{\theta(t)+\mu+\fsa\eta} \, ,
\label{eq:qss_fg}
\ee
gradually decreasing with time following $\theta(t)$.
Directly from \equ{qss_Ms}, we find that the stars-to-accreted-baryon fraction
is constant in time,
\be
\fst=\frac{\mu+\fsa\eta}{\mu+\eta} \, ,
\label{eq:qss_fst}
\ee
and so is $\stv=\p\fb\fst$.
Finally, using
\equ{qss_Mdotsf} and \equ{qss_Ms}, we find that
the sSFR is proportional to the specific accretion rate,
\be
{\rm sSFR} = \frac{\Mdotsf}{\Ms}
= \frac{\fga}{\mu+\fsa\eta}\, \frac{\Mdota(t)}{\Ma(t)}\, ,
\label{eq:qss_ssfr}
\ee
namely, based on \equ{Mdota},
\be
{\rm sSFR} = \frac{\fga}{\mu+\fsa\eta}\, s_{0.03}\, (1+z)^{5/2} \Gyr^{-1} \, .
\ee

% gass accretion only
\smallskip
For the case of gas accretion only, $\fsa=0$ ($\fga=1$), 
which could be valid at very high redshift and for relatively small masses,
the QSS expressions for the observable quantities reduce to
\be
\fg= \frac{\theta(\epsilon;t)}{\theta(\epsilon;t)+\mu}  , \quad
\fst= \frac{\mu}{\mu+\eta} , \quad
\frac{\Mdotsf}{\Ms} = \mu^{-1} \frac{\Mdota}{\Ma} .
\label{eq:qss0}
\ee
It is a very interesting feature of this model, with $\fsa=0$,
 that each of these observables is sensitive to another parameter of the model.
Assuming that $\mu$ is given, the gas fraction is determined by $\epsilon$ 
through $\theta$ and is independent of $\eta$.
In contrast, $\fst$ depends on $\eta$ only, so the observable $\stv$ 
depends on $\eta$ and $\p$.
Most interestingly, the sSFR is independent of $\epsilon$ and
$\eta$; it's average is fixed by the rather robust estimates of $\mu$ and $s$.
This is a powerful prediction of this model.

\smallskip
\Fig{mgmst} shows in dashed lines the QSS solution, 
in comparison with the exact solution, for the same choices of parameters.

%%%%%%%%%%%%%%%%%%%%%%%%%%%%%%
\section{Validity of the QSS Solution}
\label{sec:validity}

\subsection{Error of the QSS Solution}
\label{sec:error}

% validity 3
\smallskip
We can estimate the error made by this approximation at time $t$ as follows.
The time derivative of the approximate $\Mg(t)$ from \equ{qss}
is $\Mdotg=d(A\tau)/dt$, instead of the $\Mdotg=0$ obtained when
$\Mg$ from \equ{ss} is inserted in the right-hand side of \equ{equation}.
Based on this inconsistency, the error in $\Mdotg$ can be estimated
by $\Delta \Mdotg= {d(A\tau)}/{dt}$.
Then, from \equ{equation}, with $A$ given, the error in $\Mg$ is 
$\Delta \Mg = -\tau \Delta \Mdotg$.
Dividing this by the approximate $\Mg$ from \equ{qss} one obtains
\be
\frac{\Delta \Mg}{\Mg} \simeq 
-\left( \frac{\dot A}{A}\tau + \dot{\tau} \right) \, .
\label{eq:error}
\ee
The two terms correspond to the ratios of $\tau$ and the timescales for
variation in $A$ and in $\tau$, respectively, which are the two quantities that
were assumed to be small for the validity of the QSS.

\smallskip
For the time dependences of $A(t)$ and $\tau(t)$ assumed in our modeling
above,
we have from \equ{sfr} 
\be
\tau = \frac{\tsf}{(\mu + \eta)} 
= \frac{\epsilon^{-1} \nu}{(\mu + \eta)} \, t \, ,
\label{eq:tau}
\ee 
namely $\tau \propto t$ and $\dot{\tau}=\tau/t$.
From \equ{Mdota} and \equ{Ma} we derive
\be
\frac{\dot A}{A}\, t = \frac{2}{3} \alpha (1+z) - \frac{5}{3} \, . 
\ee
These give in \equ{error}
\bea
\frac{\Delta \Mg}{\Mg} 
&\!\!\! \simeq \!\!\!& - \frac{2}{3}\frac{\tau}{t} [\alpha(1+z)-1] \nonumber \\ 
&\!\!\! \simeq \!\!\!&
- \frac{0.236}{\epsilon_{0.02}\, (\mu+\eta)}\, [0.79\, (1+z) - 1] \, .
\label{eq:rerror}
\eea
For example, with $\epsilon=0.02$ and $\eta=2$, the error in $\Mg$
is 5\%, 13\%, 27\%, 57\% 
at $z=1,2,4,8$ respectively. 
%$\eta=3$ is 3.8\%, 9.1\%, 19\%, 41\% 
%
Thus, the QSS solution is a better approximation at later times.
This is despite the fact that according to \equ{tau}
$\tau/t$ is the same at all times.
The improvement with time is due to the fact that the timescale 
for variation of $\Mdota$ with respect to $t$ or $\tau$ 
[namely $(A/\dot{A})/\tau$] becomes longer in time.
Note that the error associated with the QSS solution depends on all the
model parameters, and, in particular, it is a decreasing function of $\eta$.

\smallskip
The relative errors in the SFR, $\Ms$ and $\Mb$ are expected to be
comparable to the relative error in $\Mg$.
The error in $\fgt$, $\fst$ and $\fbt$ are therefore all expected to be
comparable, and estimated by \equ{rerror}, because $\Ma$ is given with no 
error. 
On the other hand, the errors in $\fg$ and $\fs$, as well as in the sSFR,
are expected to be of higher order and much smaller, 
because the errors in the numerators and in the denominators are similar 
and correlated. 

\smallskip
\Fig{mgmst} shows that, once the transient components decay, 
the exact solution also approaches a unique asymptotic solution, independent
of the initial conditions.
The exact solution for $\fst$ deviates from the approximate QSS solution due to 
the time variation of $A$ (and $\tau$) following the error estimate in 
\equ{rerror}. This deviation gradually diminishes at later times.
\Fig{mgmst} (right) shows how this deviation depends on $\eta$ -- it is 
at the level of less than 25\% for $\eta=3$, 25-40\% for $\eta=1$,
and a factor of 2 or more for $\eta=0$.
Even better than expected, the QSS solution for $\fg$ and for the sSFR 
almost coincides with the exact asymptotic solution, making the analytic
QSS solution very powerful.

%-------------------------
\subsection{Decay of Transients}
\label{sec:transient}

% validity 2
The decay of the transient component in the exact solution can be estimated
from its decay in the SS solution, \equ{ts}, namely $\propto \exp(-t/\tau)$.
According to \equ{tau}, $t/\tau$ is time-independent,
\be 
\frac{t}{\tau} \simeq 2.82\, \epsilon_{0.02}\, (\mu+\eta) \, .
\label{eq:tovertau}
\ee 
Thus, the asymptotic regime $t \gg \tau$ is valid at all times
as long as $\epsilon$ and $\eta$ are sufficiently large.
An exponential decay of the transient component 
by an order of magnitude, namely $t/\tau\simeq 2.3$, requires 
$\mu+\eta \geq 0.82\, \epsilon_{0.02}^{-1}$.
For $\epsilon \simeq 0.02$ this condition is obeyed for any sensible value of
$\eta$, while for $\epsilon \simeq 0.01$ it requires that $\eta$ exceeds
unity.
Note that the validity of the SS solution, $t\gg \tau$, depends on $\epsilon$
and $\eta$, but is independent of the accretion rate $A$ (namely 
$\fga \Mdota$).

\smallskip
As explained in \se{qss_solution}, if the initial conditions significantly
deviate from the QSS solution, the corresponding transients in the solution
for $\Ms$ are expected to decay as $\exp(-\alpha \Delta z)$. 
This is demonstrated in \fig{mgmst}.
Shown in the left panel are three cases with different values of $C$ in the 
initial conditions at $\zi=10$, specifically $C/\Mai = -0.25,0,+0.5$. 
We see that the exact solution converges to its unique asymptotic solution by 
$z \!\sim\! 6-7$, i.e., the transients decay significantly during 
$\Delta z \!\sim\! 3$, as expected.

%If, somehow, $\tau$ is constant in time, on the order of 1 Gyr,
%the approximation would be valid only after $z \!\sim\! 5-6$.

%--------------------
\subsection{A Physical QSS Solution}
\label{sec:physical}

\smallskip
\Equ{qss_Mg} implies
\be
\fgt = \frac{\theta}{\mu+\eta} \, .
\label{eq:fgt}
\ee
Since $\theta$ is decreasing with time, \equ{theta},
$\fgt$ must also be decreasing with time.
The by-definition requirement that $\fgt \leq 1$ induces a
constraint on $\theta(t)$ for a given $\eta$,
\be
\theta(t) \leq \mu+\eta\, .
\label{eq:validity1}
\ee
Since $\theta$ is decreasing with $t$, 
this translates to a lower limit on the time where the
QSS solution could provide a physical solution with $\fgt \leq 1$. 
The interpretation of this constraint is that prior to this time
the available gas mass is insufficient for the SFR to catch up with the intense
accretion rate. During this early period, the gas mass grows in time
and the accretion
rate gradually declines, until the SFR can catch up with the accretion rate
and the QSS solution is approached.
Using \equ{theta}, with $\epsilon=0.02$, we see 
that for $\eta=1$ the QSS solution is physical 
in the range $z\leq 7.3$, and for $\eta \geq 1.5$ it is physical for
$z \leq 9.7$, namely in the whole relevant range. 
However, for $\epsilon = 0.01$, with $\eta=1$ the physical solution is limited
to $z\leq 4.2$, but with $\eta=3$ it is valid in the whole range, $z\leq 8.5$.
Note that this validity criterion depends on the parameters
$\epsilon$ and $\eta$ as well as on $\fga$ and $\Mdota$.

%\adr{xxx Possible figure for errors}

%2
\begin{figure*}
\vskip 7.1cm
\includegraphics{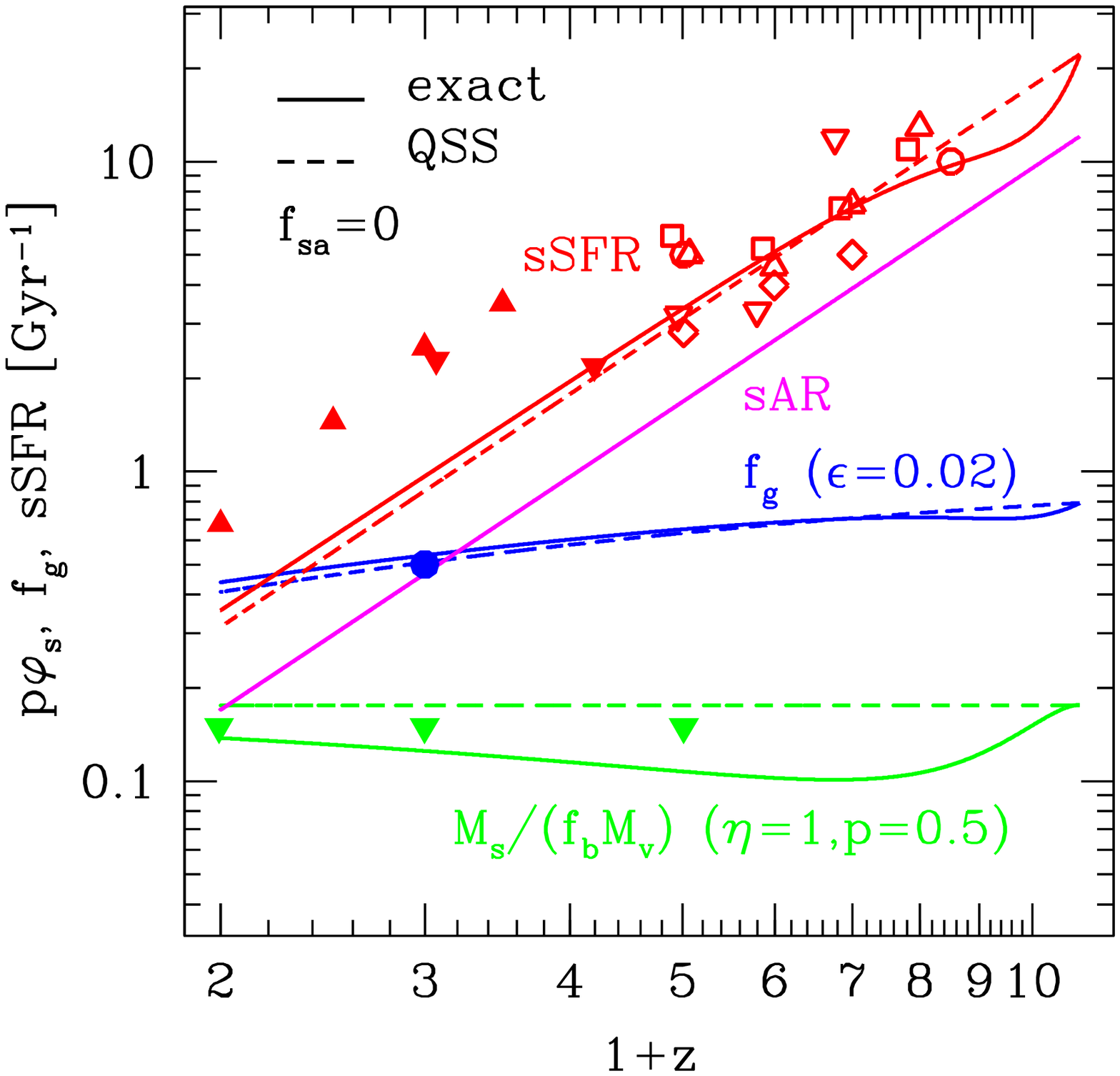}
\includegraphics{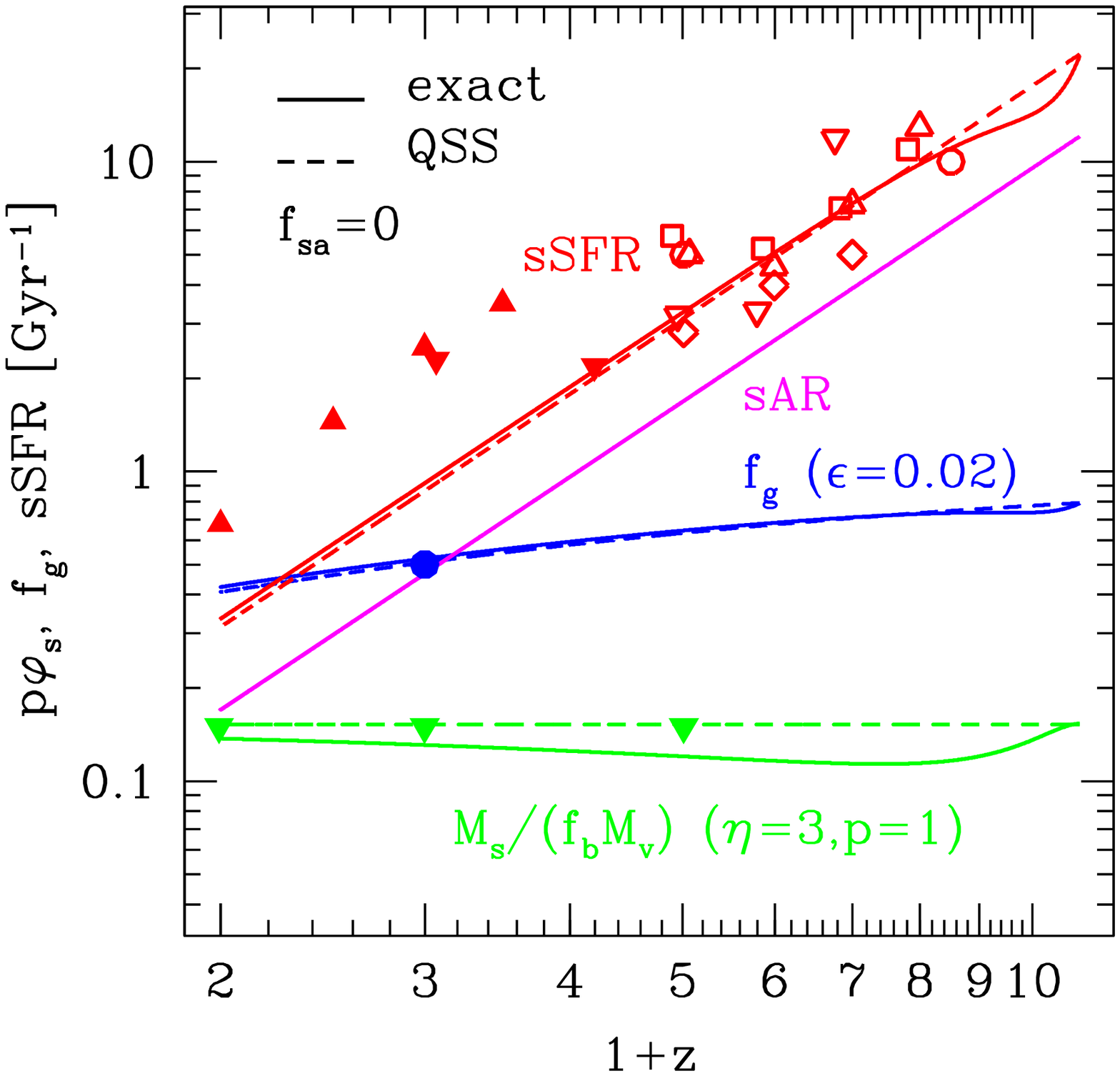}
\caption{
Solutions of the bathtub model with gas accretion only, $\fsa=0$,
with the average sAR, compared to observations.
For more details see \fig{mgmst}.
Observational indications are marked by symbols.
The sSFR measurements (red) are from
\citet{whitaker12_ssfr} (filled triangles),
\citet{reddy12} (filled up-side-down triangles),
\citet{gonzalez12} (up-side-down triangles),
\citet{labbe13} (circles),
\citet{stark13} (squares),
\citet{duncan14} (triangles),
\citet{salmon14} (diamonds).
The gas fraction $\fg$ at $z\sim\! 2$ is from \citet{tacconi13} (blue circle).
The stellar-to-virial mass ratio
$\stv/\fb$ at $z=1-4$ is based on \citet{behroozi13} (green triangles).
This bathtub model robustly reproduces the sSFR at $z=3-8$,
with $\epsilon=0.02$ uniquely constrained for matching the observed $\fg$.
The value of $\eta$ is constrained to be $\eta=1$ (left) or $\eta=3$ (right)
for $\p\fst$ to match the observed $\stv/\fb$ with a penetration factor
of either $\p=0.5$ or $\p=1$ respectively.
The only exception is the sSFR at $z \!\sim\! 2$, which is under-predicted by a
factor of $\sim\! 2\!-\!3$.
}
\label{fig:mgms0}
\end{figure*}

%%%%%%%%%%%%%%%%%%%%
\section{Comparison to Observations}
\label{sec:comp_obs}

\subsection{Observations}
\label{eq:obs}

\smallskip
We address here three rather intriguing observational results at high 
redshifts, indicated as symbols in \fig{mgms0}.

\smallskip
First is the average sSFR for massive galaxies as a function of redshift.
This either applies to galaxies selected to have a fixed mass at different
redshifts, or to have the same comoving number density, thus
assumed to mimic an evolving sample of galaxies.
At $z \leq 2$, the results are relatively reliable thanks to measurements 
of H$_\alpha$ and deep far-IR data.  The average sSFR declines in time from 
$2-2.5 \Gyr^{-1}$ at $z = 2$, through $\simeq 0.7 \Gyr^{-1}$ at $z = 1$,
toward $\sim\! 0.1-0.2 \Gyr^{-1}$ at $z=0$ \citep{whitaker12_ssfr,reddy12}.
The high sSFR at $z \!\sim\! 2$  
%D
and its potential conflict with theory have been noticed \citep{daddi07},
and led to considerations of a top-heavy IMF as a possible remedy
\citep{dave08}.

\smallskip
At higher redshifts, the estimate of sSFR requires non-negligible modeling
and assumptions. The uncertainties in the stellar population models,
and especially the tentative implementation of emission lines, 
makes the results subject to systematic uncertainties by a factor of 2 or more.
Early results indicated an apparent ``sSFR plateau", with a constant sSFR 
in the range $z=2-8$ \citep[e.g.][]{stark09,gonzalez10,labbe10}.
Attempts to model this sSFR plateau revealed severe difficulties
\citep{weinmann11}, and \citet{kd12} showed that it requires a non-negligible 
fraction of stars in the accretion.
However, 
recent estimates have corrected the observed behavior of the sSFR at high
redshifts, now showing a continuous decline with time. 
The calibration of absolute levels is more uncertain, with estimates of 
sSFR$\sim\! 10 \Gyr^{-1}$ at $z=7-8$ declining to $\sim\! 2-3 \Gyr^{-1}$ 
at $z\!\sim\!3$ 
\citep{gonzalez12,stark13,labbe13,duncan14}.
In particular,
estimates based on SED-fitting, that incorporates effects
of nebular line emission and considers both starburst and SMC-type
dust attenuation curves, reveal absolute values on the lower side 
\citep{salmon14}.
It is encouraging that the decline rate is not very different from the 
decline predicted for the 
specific accretion rate, but the fact that the observed sSFR is still at a 
somewhat higher amplitude is challenging.

\smallskip
The second observation is the gas fraction in the galaxy,
which is deduced from CO observations
to be $\sim\! 0.5$ at $z \!\sim\! 2$ and declining with time
\citep{daddi10,tacconi10,tacconi13}. Many simulations fail to
reproduce such high gas fractions as late as $z \!\sim\! 2$, as star formation
in these simulations tends to consume most of the gas earlier.
%D
However, systematic errors are possible due to the assumed CO-to-gas conversion
\citep{narayanan12}
and due to a potential selection bias toward gas-rich galaxies
\citep{tacconi13}.

\smallskip
The third observation is the stellar-to-virial mass ratio, estimated from
observations of stellar masses using abundance matching of dark-matter haloes 
from simulations \citep{guo10,moster10,moster13,behroozi13}.
As can be seen in Figure 3 of \citet{behroozi13}, 
it turns out to be rather independent of redshift in the range $z=0-4$,
with a maximum value of $\stv\!\sim\! 0.025$, obtained for halo masses of
$\Mv \!\sim\! (0.5-5)\times 10^{12}\msun$. Given the universal baryon fraction
$\fb \simeq 0.17$, this implies that only $\sim\!15\%$ of the baryons that
were supposed to enter the halo have made it into stars at the central galaxy,
namely $\stv/\fb = \p\fst \simeq 0.15$,
which is a non-trivial challenge for the modeling.
We tentatively adopt this low estimate, but one should mention that
new determinations of stellar masses in bright central galaxies of
clusters indicate that the stellar-to-virial ratio for very massive haloes
at low redshift is actually higher then the earlier estimates, 
by a factor of 2-4 \citep{bernardi13,krav14}. 
This raises the potential worry that the adopted $\stv$ is an underestimate
also on the scales of interest here, $\sim\!10^{12}\msun$, 
and at high redshifts. 

\smallskip
Significant outflows are observed in high-redshift galaxies
\citep{steidel10,genzel11}, with a mass-loading factor ranging from below unity
to 10, but the large uncertainty involved in these estimates makes
us leave it as an open variable, to be determined by model fitting to the other
observations.
The situation is similar regarding the observed SFR efficiency
\citep{krum_tan07,krum_thomp07},
which we leave free to be constrained by the other observations.

%3
\begin{figure*}
\vskip 7.1cm
\includegraphics{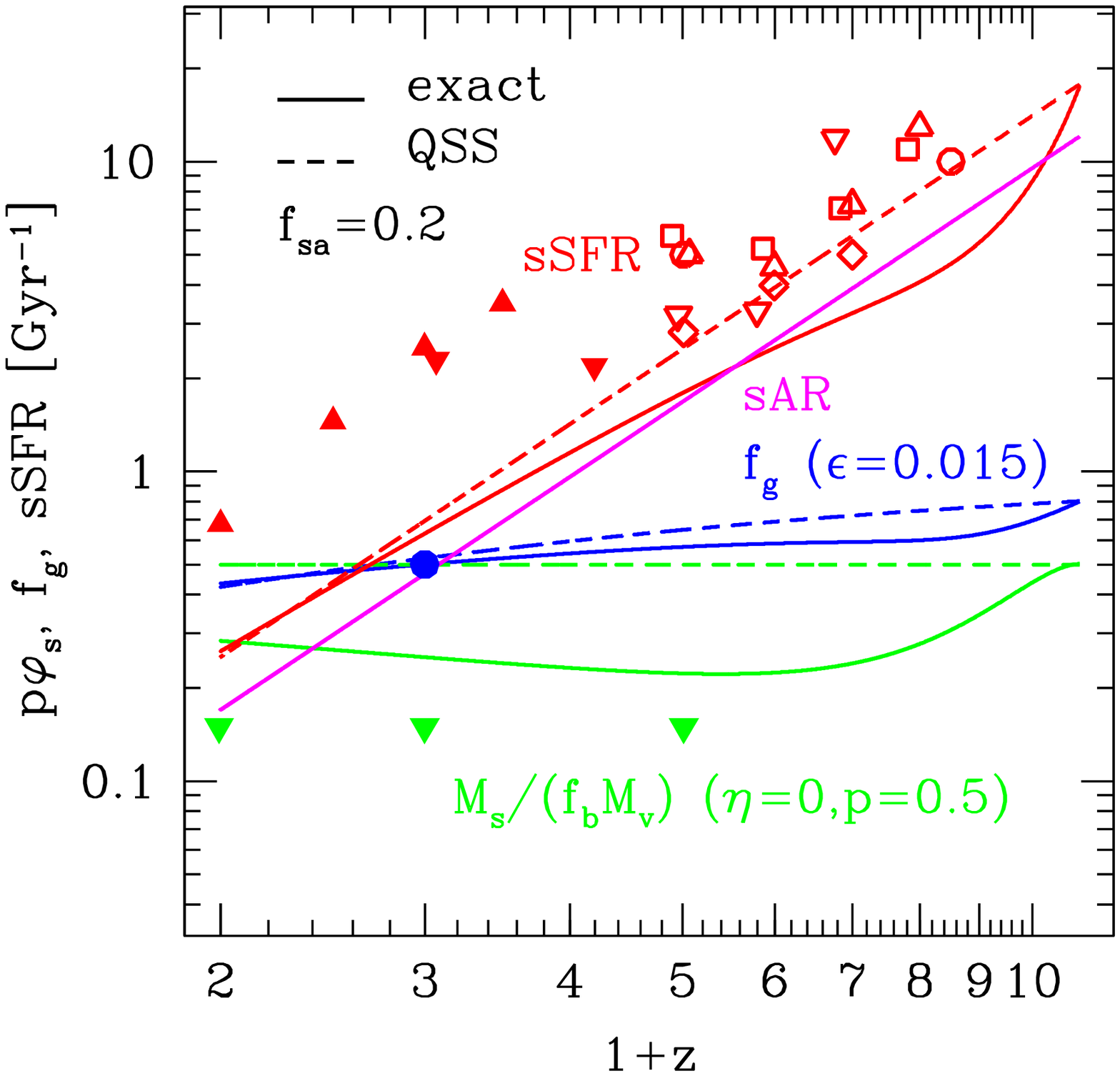}
\includegraphics{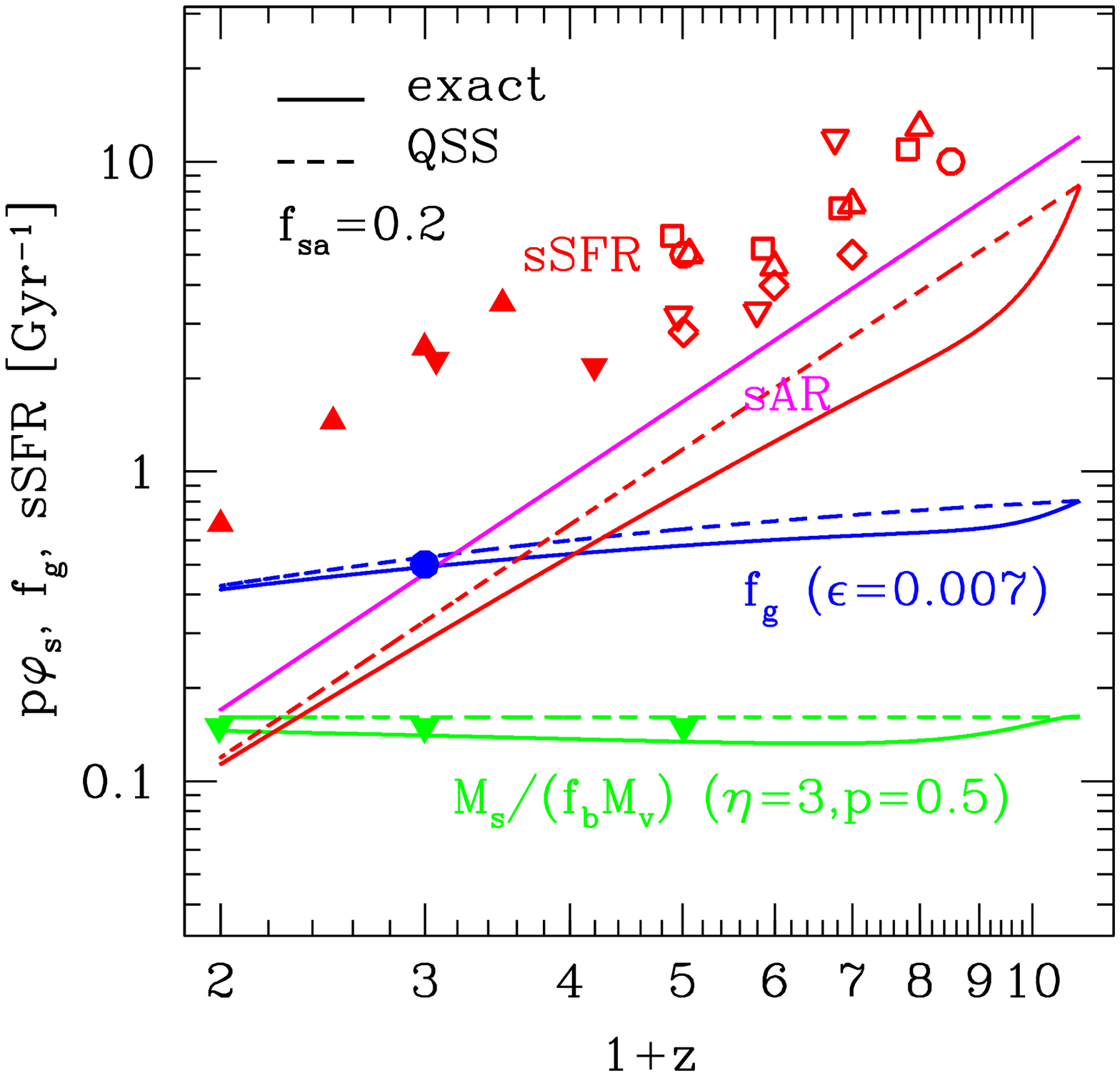}
\caption{
Solutions of the bathtub model including stellar accretion, $\fsa=0.2$,
with the average sAR, compared to observations.
For more details see \fig{mgmst} and \fig{mgms0}.
{\bf Left:} With maximum-recycling $\eta=0$ for maximum sSFR, 
and the best-fit $\epsilon=0.015$ for $\fg$. 
Even with this low $\eta$ the sSFR is slightly under-predicted
at $z=3-8$, and by a factor of $\sim\! 3$ at $z=2$, while
the $\stv/\fb$ is over-predicted by $\sim\! 50\%$ even with $\p=0.5$.
{\bf Right:} With $\eta=3$ for a match of the $\stv/\fb$ with
$\p=0.5$, and the best-fit $\epsilon=0.007$ for $\fg$.
Now the sSFR is under-predicted by a factor of $\sim\! 3$ at $z=3-8$,
and by a factor of $\sim\! 7$ at $z\sim\! 2$.
}
\label{fig:mgms_fga0.2}
\end{figure*}

%--------------------
\subsection{Gas Accretion Only}
\label{sec:fga}

% QSS
The simple QSS solution for the case of gas accretion only, $\fsa=0$, is 
given in \equ{qss0}. We can use it to evaluate the fit to the observed sSFR
and estimate the best-fit values for $\epsilon$ and $\eta$, 
which can then be fine-tuned
using the exact solution. 
\Fig{mgms0} shows the observational constraints in comparison to 
the QSS and exact solutions with the best-fit choice of parameter values, as
follows.

\smallskip
In order to obtain $\fg \!\sim\! 0.5$ at $z \simeq 2$, 
one needs in \equ{qss0} a value of $\theta \simeq 0.5$, which, 
based on \equ{theta}, is obtained at $z \simeq 2$ for $\epsilon \simeq 0.02$. 
Then, the value of $\fg$ is predicted to be gradually larger at higher 
redshift, e.g., $\fg \!\sim\! 0.8$ at $z \!\sim\! 10$.  
A value of $\epsilon \simeq 0.01$ gives at $z=2$ a gass fraction of 
$\fg \simeq 0.67$, 
which is too high, while $\epsilon \simeq 0.04$ gives $\fg \simeq 0.34$, 
which is too low.

\smallskip
For the comparison with $\stv$ we assume a penetration of $\p=0.5$.
In order to obtain the required low value for $\fst$ in \equ{qss0}, one needs
non-negligible outflows, $\eta \!\sim\! 1$.
With $\p=1$, stronger outflows are required, $\eta \!\sim\! 3$.
The solution for $\p\fst$ is shown in \fig{mgms0} for $\p=0.5$ and for $\p=1$.
The fact that the exact solution for $\fst$ is lower than
the corresponding QSS solution makes it easier to match the low observed
$\stv$ with somewhat lower values of $\eta$. 

\smallskip
Finally, with $\mu \simeq 0.54$, the sSFR is roughly twice the specific 
accretion rate at all times in the QSS regime, independent of $\epsilon$ or
$\eta$. There is no choice of the model
parameters that can give on average higher values of sSFR.
The predicted values are sSFR=$0.94,3.4,7.8,14.6 \Gyr^{-1}$ at $z=2,4,6,8$ 
respectively.  This is in good agreement with the observations at $z=3-8$.
However, at $z \!\sim\! 2$, the predicted average sSFR falls short by a factor of 
$\sim\! 2$.
We will return to possible solutions to this puzzle in \se{disc_z2}.

%--------------------
\subsection{Including Stellar Accretion}
\label{sec:fsa}

With a non-vanishing value for the stellar fraction in the accretion, $\fsa$, 
the model faces difficulties in reproducing the observations.
This is because the stellar accretion makes $\Ms$ grow without a corresponding
growth in SFR, thus pushing the sSFR down and $\stv$ up 
(e.g., see the effect of a non-negligible $\fsa\eta$ in
the QSS solution, \equ{qss_ssfr} and \equ{qss_fst} respectively). 
Furthermore, with $\fsa > 0$, each of the observables becomes sensitive 
to more parameters than for $\fsa=0$, 
and in particular the sSFR is pushed down, more so when $\eta$ is larger.
Also, the deviations of the QSS toy solution from the exact solution become
larger, especially for the sSFR.

\smallskip
\Fig{mgms_fga0.2} demonstrates these difficulties in the case $\fsa=0.2$.
The left panel refers to a low value of $\eta=0$, trying to push the sSFR up
as much as possible.
A value of $\eta \simeq 0$ may in fact have a sensible physical interpretation,
where it represents a balance between outflow and recycling.  
A value of $\epsilon=0.015$ is now required for a good match to $\fg$ at 
$z\sim\! 2$, with no strong effect on the matching of the other observables.
While the QSS solution is now a good match to the sSFR at $z=3-8$,
the exact solution slightly under-predicts the sSFR at $z=3-8$, 
by $\sim\! 30\%$, 
and the predicted sSFR falls short more severely at $z\sim\! 2$, by a factor of 
$\sim\! 3$. 
With this low $\eta$, the exact solution for $\stv$   
overestimates the observed value by $\sim\! 50\%$ even with $\p=0.5$. 
If $\p=1$, the deviation from the observed value becomes a factor of 
$\sim\! 3$.  One needs $\p \!\sim\! 0.3$ for a match.
Note that the lowest possible value of $\fst$ is $\fsa$, 
obtained when $\eta$ grows to infinity. Thus, a value of $\fsa>0.15/\p$
will make it impossible to match the observed $\Ms/(\fb\Mv)=0.15$.

\smallskip
The right panel of \fig{mgms_fga0.2} appeals to stronger outflows, $\eta=3$, 
as required for a match of $\stv$ with $\p=0.5$.  For this $\eta$,
a value of $\epsilon=0.007$ provides the desired match to $\fg$ at $z\sim\! 2$.
However, 
now the failure in matching the sSFR becomes more severe -- an under-prediction 
by a factor of $\sim\! 3$ at $z=3-8$, and by a factor of $\sim\! 6-7$ at 
$z\sim\! 2$

%%%%%%%%%%%%%%%%
\section{Discussion}
\label{sec:disc}

\subsection{Model Fits Observations at $z = 3-8$}
\label{sec:disc_z38}

We should first recall that the asymptotic solution of the bathtub model 
with a given set of parameter values can be considered at a given cosmological 
period independent of whether it was valid at an earlier time.

% fsa=0
\smallskip
At sufficiently high redshifts, and more so for less massive galaxies,
the approximation of gas accretion only, $\fsa=0$, may be naturally acceptable.
In this case, the simplest possible bathtub toy model turned out to be 
extremely successful in qualitatively reproducing the observations, 
and in constraining the rates of star formation and outflows.
The QSS solution naturally reproduces the constancy in time of the 
stellar-to-virial mass ratio, $\propto\!\fst$, 
and the gradual decrease of the sSFR 
with time at a level comparable and slightly higher than the theoretically 
estimated specific accretion rate.
The observational requirement of $\fg \!\sim\! 0.5$ soon after $z \!\sim\! 3$
dictates an SFR efficiency in a disk dynamical time of 
$\epsilon \!\sim\! 0.02$, 
which is consistent with what we knew a priori
\citep{krum_tan07,daddi10_sfr,genzel10}.
Independently, the requirement of $\p\fst \!\sim\! 0.15$ dictates a 
mass-loading factor $\eta \!\sim\! 1$ if the penetration is $\p \!\sim\! 0.5$, 
as deduced from simulations (and $\eta \!\sim\! 3$ if $\p \!\sim\! 1$).
This is consistent with the observational indications for rather intense 
outflows \citep{steidel10,genzel11}.
This toy model, and its analytic QSS solution, can thus be very useful in the 
interpretation of observations at very high redshifts, say $z \geq 3$.  

% fsa=0.2
\smallskip
However, at $z \!\sim\! 2$, and in massive galaxies, the stellar accretion via 
mergers is not likely to be negligible \citep{dekel13}.
Even for $\fsa \!\sim\! 0.2$, and especially with non-negligible $\eta$,
this component dominates the growth of stellar mass.
It drives $\Ms$ up independent of the SFR, and thus pushes the sSFR down
and $\fst$ up, away from the observed values.
In this case, the model cannot simultaneously provide a perfect match
to the high sSFR and the low $\fst$, as the former favors vanishing effective
outflows, $\eta \ll 1$, while the latter requires intense outflows,
$\eta \!\sim\! 3$. The case $\eta=3$ is unacceptable, because the sSFR is 
severely under-predicted, by a factor of 3 at $z=3-8$ and by a factor of 7 at 
$z\sim\! 2$.
The case $\eta=0$, on the other hand, provides a potentially acceptable
compromise, where the sSFR at $z=3-8$ is under-predicted by only 30\% and  
$\stv$ is over-predicted by 50\% for $\p=0.5$.
The latter is properly matched if $\p=0.3$.
Such a low value of $\p$ is possible 
since the haloes at $z \!\sim\! 2$ and later are more likely to become more 
massive than the critical mass for virial shock heating \citep{db06}, 
where the penetration into the galaxy becomes more difficult.
An alternative solution arises if the adopted low observed value of $\stv$ 
is an underestimate \citep[as might be indicated by][]{krav14}. 
An increase of $\sim\!50\%$ in the observed $\stv$ 
is needed for the model with $\eta=0$ to match it.

%-------------------------
\subsection{A Challenge at $z \!\sim\! 2$}
\label{sec:disc_z2}

% z=2 challenge
\smallskip
The bathtub model highlights the fact that the sSFR at $z \!\sim\! 2$ 
introduces 
a non-trivial challenge, with the observed value being $\sim\! 3$ times the 
average specific accretion rate.
Even for gas-accretion only, the predicted sSFR is short by 
a factor $\sim\! 2$ compared to the observational estimates.
With a relatively small component of stellar accretion, $\fsa=0.2$,
and with $\eta=0$, the deviation becomes a factor of $\sim\! 3$.
Higher values of $\fsa$ or $\eta$ would make this deviation more severe.
Thus, matching the high sSFR at $z \!\sim\! 2$ is not easy.
We address possible solutions next.

%-----
\subsubsection{Recycling}
\label{sec:recycling}

%Recycling: 
A potential explanation for the high sSFR at $z\sim\! 2$ is recycling,
the return of gas that has been ejected earlier by feedback but remained
bound. 
Since gas is likely to accumulate in the galaxy at earlier periods when the 
SFR is low compared to the accretion rate,
it may be available for strong outflows (especially in low-mass galaxies),
to be followed by recycling into more massive galaxies at $z\!\sim\! 2$.
The low SFR at very high redshifts may be due to the effect discussed in
\se{physical} where more gas is needed for the SFR to catch up with the intense
accretion rate. It may also arise from a low SFR efficiency parameter
$\epsilon$, e.g., due to low metallicity in the galaxy building
blocks \citep{kd12}.

\smallskip
Recycling is modeled in our minimal bathtub model in a very crude way, 
as a negative contribution to $\eta$. The model with $\eta=0$ may thus be 
regarded as representing a period of intense recycling, where the return 
is at a rate comparable to the outflow rate at that time. 
With $\fsa=0.2$, this brings the underprediction of the sSFR at 
$z \!\sim\! 2$ to the level of a factor of $\sim\! 3$.
%this brings the discrepancy at $z \!\sim\! 2$ to a factor of $\sim\! 3$.
%Furthermore, the recycled gas increases the gas fraction in the accretion, 
%and if the recycling dominates, it may make the stellar accretion negligible
%such that the model with $\fsa=0$ is valid. This brings the deviation of the
%sSFR at $z \!\sim\! 2$ to the level of a factor of $\sim\! 2$, as in \fig{mgms0}.
%
The remaining deviation may possibly be reduced with a more sophisticated
implementation of recycling in the modeling.

%------
\subsubsection{Other Parameters}
\label{sec:other}

\smallskip
One possibility for easing the tension concerning the sSFR at 
$z \!\sim\! 2$ is to consider
different average values for one or more of the model parameters. 
For example, the value of $s$ describing the average sAR in \equ{Mdota}, 
deduced from simulations \citep[e.g.][]{dekel13}, 
may possibly be an underestimate because
the sample of galaxies simulated is biased against high-density environments,
which are indeed more abundant at $z \!\sim\! 2$ than earlier. 
A higher value of $s$ would reproduce a higher sSFR, as desired.

\smallskip
Another possibility is that $\mu$, the fraction of 
star-forming mass that remains in stars, may actually be lower as a result 
of a top-heavy IMF. In this case one may wonder why this would happen at 
$z \!\sim\! 2$ and not earlier.

%-------
\subsubsection{Bias due to quenched galaxies}
\label{sec:bias}

% general idea. correlation of low model sSFR with quenching
The high average sSFR indicated by observations at $z \!\sim\! 2$ 
may reflect an observational selection bias toward the high sSFR galaxies,
in the presence of scatter in the sSFR of different galaxies.
Recall that galaxies are selected from the SFG population,
which, by $z \!\sim\! 2$, constitute only about half the massive galaxies,
the rest being already quenched by that time 
\citep[e.g.][]{kriek06,dokkum08,tal14}. 
A scatter can be introduced in the parameters of the bathtub model, 
that would lead to scatter in the predicted sSFR at a given time.
However, our toy model does not attempt to model the actual quenching process,
so the low-SFR galaxies are still star forming.
The key for this effect to lead to a better agreement with the observed
sSFR is that the sources for relatively low sSFR in the model would actually 
lead to an even lower sSFR, namely quenching, once quenching is incorporated,
as in the real-Universe galaxies.

% scatter in parameters
\smallskip
According to \equ{qss_ssfr},
low sSFR is obtained in the model if $s$ that characterizes the specific
accretion rate in \equ{Mdota} is low,
and if $\mu+\fsa\eta$ is high, namely when either $\mu$ or $\fsa\eta$ or both
are high.  We consider each of these sources of scatter.
According to cosmological simulations, the distribution of $s$ among snapshots, 
including scatter between galaxies and along the evolution of each galaxy,
is $\pm 0.45$dex \citep[][Figure 7]{dekel13}.
Based on hydro-cosmological simulations with supernova and radiative stellar
feedback, we deduce a typical scatter of $\pm 0.3$dex in $\eta$ and 
a similar scatter % $\pm xxx$ xxx  check Tweed
in $\fsa$ among different output snapshots 
(House et al., in preparation; Tweed et al. in preparation).
Scatter in $\mu$ may reflect variations in the IMF.

% correlation with quenching
\smallskip
High values of $\eta$ and $\fsa$ may indeed lead to quenching. 
A high $\eta$ indicates gas removal by stellar or AGN feedback
(with little recycling) that would naturally lead to quenching.
A high $\fsa$ is associated with a high dry-merger rate, 
which tends to occur in galaxy cluster environments, namely in more massive 
dark halos. Such halos are likely to be more massive than the critical mass for
virial shock heating, $\sim\! 10^{12}\msun$, and thus likely to be subject
to halo-mass quenching \citep{db06,woo13,db14}. 
A low sSFR could occur in particular in a galaxy that is observed in a quiet 
accretion period but had significant gas-poor merging in the recent past, 
such that $\Ms$ grew significantly but the SFR is similar to what it was
before the mergers. In such a post-merger galaxy, which could also be
post-starburst and post-outflow, quenching follows by gas consumption and
removal.   
As for low values of $s$ or high values of $\mu$, while it is clear that they
lead to low sSFR in the model, it is not clear whether they would lead
to quenching in real galaxies. 

%--------
\subsubsection{Breakdown of the QSS}

Another possible solution to the discrepancy at $z \!\sim\! 2$ is a breakdown of
the validity of the QSS solution over a short period of cosmological time. 
This could result from strong fluctuations in the gas accretion rate $A$
with a short timescale, or similar fluctuations in the characteristic 
timescale $\tau$. Such fluctuations may invalidate the QSS
and limit the usefulness of the bathtub model for describing the 
instantaneous properties of galaxies, while it may still
trace the long-term average properties. 
For example, the sSFR may get temporarily high without a corresponding
increase in $\Ms$ as a result of a star burst or a recycling burst.

%---------------------------
\subsection{Validity of the Model Assumptions}
\label{sec:disc_assumptions}

The ingredients of the minimal bathtub toy model discussed here
involve certain simplifying assumptions worth discussing concerning 
the cosmology, accretion rate, SFR and outflows.

% to z=0
We have applied the model in the EdS regime, $z >1$, in order to allow a simple
analytic solution. However, the model can be easily applied in the regime 
where the $\Lambda$CDM cosmology deviates from the EdS approximation, down
to $z=0$. In this regime, the accretion rate deviates form the simple form of
\equ{Mdota}, so the QSS solution has to be derived numerically.
An extrapolation of the EdS expression to $z=0$ gives a value of sSFR below 
$0.1\Gyr^{-1}$, already qualitatively consistent with observations, 
so it recovers much of the low-redshift decline without appealing to the
acceleration of the Universe or to baryon-physics processes. 
Note that when plotting the sSFR as a function of $z$ rather than $\log(1+z)$,
the decline of the sSFR shows the familiar steepens with time after 
$z \!\sim\! 2$, following the associated drop in the sAR.
The decline gets steeper when taking into account the 
acceleration of the Universe at $z<1$, and the quenching of massive galaxies
in shock-heated haloes at low redshifts \citep{db06,vandevoort11}. 
An extrapolation of the EdS solution to $z=0$ also yields a gas fraction
estimate of $\fg \sim 0.1$ (for $\eta\!\sim\!1$ and $\fsa\!\sim\!0.5$, say), 
and a value of $\stv$ similar to its value at higher redshifts, 
both in the ballpark of the average observed values. 
The solution taking into account the acceleration phase is not very different.

% instantaneous
A general limitation of the bathtub model is that 
the continuity equations refer to a given time, and the ingredients are all
assumed to be instantaneous. In reality, the star formation lags behind the
accretion of the gas involved, and the mass loss from stars ($\mu$), the
outflows from the galaxy ($\etaout$), and especially the recycling 
($\etarec$), all occur after the relevant events of star formation. 
The instantaneous quantities are sensible approximations once the system 
is indeed in or close to a quasi-steady state, 
and where the time delays are shorter than the timescale for long-term 
variations in quantities such as the accretion rate and the star-formation 
time.

% mass-independent
\smallskip
In the simple version of the toy model addressed here, 
mass dependence enters only through
the linear mass dependence of the accretion rate and the SFR (virial mass and
gas mass respectively).
The model neglects any mass dependence of the specific
accretion rate, $\tsf$, or $\eta$, as well as $\mu$ (and $\p$).
This is a reasonable approximation over a certain mass range, 
but it may fail outside this range.
%D
A mass dependence may have certian consequences, e.g.,  
a suppressed sSFR in low-mass galaxies at high redshifts would tend to
increase the amount of gas available for star formation at $z\!\sim\!2$
\citep[e.g.][]{bouche10,kd12}.
%D
Introducing a mass dependence is possible, but it requires numerical
integration of the bathtub model for a population of galaxies of different 
masses, and convolving the results with the mass function of galaxies. 
By integrating the continuity equations of the bathtub model
one follows the time evolution of a given galaxy as it grows.
This can be naturally compared to the average over 
observed galaxies selected at a fixed comoving number density at different 
redshifts, assumed to mimic the evolution of a given sample of galaxies,
ignoring mergers. 
In the absence of mass dependence in the sAR, $\tsf$ and $\eta$,
the predictions can also be compared to galaxies selected to have the same 
mass at the different redshifts.
Indeed, the observed sSFR evolution for samples of galaxies selected 
in those two different ways is rather similar \citep[][Figure 17]{salmon14},
indicating mass independence 
%D
in the relevant mass range.

% fsa=fs
\smallskip
The stellar fraction in the accretion turned out to be a key factor.
Given the uncertainty in this fraction,
one may be tempted to try the extreme case where it is the same as the
stellar fraction within the evolving galaxy at the same time, $\fsa=\fs$.
This turns out to generate a rapid drop in $\fg$ soon after the start of the 
integration, in sharp disagreement with observations.
We learn that $\fsa$ should grow much slower than $\fs$, consistent with the
fact that a large fraction of the accreted baryons is associated with galaxies
of much lower mass and with smooth accretion, including massive recycling. 
A more reliable estimate of $\fsa$ at different epochs is yet to be determined.

% SFR prop Mg
\smallskip
The assumption that the SFR is proportional to the total gas mass is clearly a
crude simplification, as we know that the SFR actually follows the molecular 
gas surface density, which does not necessarily follow the total gas mass
density in the low-density and low SFR regime. 
Nevertheless, it is an approximation worth adopting even if crude, in the
regime where it is not totally off, 
because it is the feature that drives the system into the self-regulated
situation and makes it simple to work out.

% tsf=const
\smallskip
We adopted above the most natural assumption that the depletion time $\tsf$ is  
proportional to the disk dynamical time, and that the latter
is proportional to the cosmological time.
There is an observational indication for $\tsf \propto (1+z)^{-1}$
out to beyond $z \sim 2$ \citep{saintonge13}, which is not far from the
$\tsf \propto t$ assumed in the toy model.
These observations are also consistent with the SFR being driven by the total
gas mass.
However, the assumption of $\tsf \propto t$ 
may break down in different ways, e.g., the depletion time
may vary in a slower pace, to the extreme that it may be close to constant 
in time in a certain epoch % \citep{xxx}. Check Burkert
At a given time, this would be like adopting a different SFR
efficiency $\epsilon$, which would mostly affect the gas fraction $\fg$,
but would have little effect on the sSFR and on $\stv$.
%\adr{check}

% eta
\smallskip
The assumption that the outflow rate is proportional to the SFR with a constant
mass-loading factor is supported by observations and is sensible theoretically 
for stellar feedback, but it is an approximation that may fail for AGN
feedback and for other mechanisms that could be driving outflows.
Our modeling of recycling as an instantaneous negative contribution to 
$\eta$ is clearly simplified, and it may or may not capture the main features
of the recycling process, which may be crucial for sorting out the discrepancy 
at $z \sim 2$.

% p
\smallskip
The penetration parameter $\p$ is not a necessary ingredient of the toy model.
It is required only for the comparison of the model with observational
estimates that involve the total halo mass, such as $\stv$.
For a crude comparison with this observation, 
the typical value $\p \sim 0.5$ 
deduced from current simulations is appropriate, 
but a more detailed estimate of $\p$ from simulations that span a range of
feedback strengths is desired.

%%%%%%%%%%%%%%%%%%
\section{Conclusion}
\label{sec:conc}

The minimal bathtub toy model is shown to be a useful tool for 
tracing the roles played by key processes of galaxy evolution
and identifying major
successes and tentative failures in reproducing observations at high redshift.
This model is based on simplified but robust continuity equations for the gas 
and for the stars.
Because of the generic monotonic dependence of the SFR (and therefore outflow
rate) on gas mass,
the system is self-regulated into a unique asymptotic behavior, which
can be approximated by a quasi-steady-state solution.
The simple time and mass dependence of the average accretion rate into
galaxies allows an approximate analytic quasi-steady state solution. 

\smallskip
We derived the analytic QSS solution and evaluated the associated deviation
from the exact solution for three observables: 
the sSFR, the gas fraction, and the stellar-to-virial mass fraction.
We showed that the error is negligible for the first two, and is limited to
the level of tens of percent for the third quantity.
We studied the range of validity of the QSS solution, and found that 
the errors are smaller at later epochs, and when the outflows are stronger.

\smallskip
In the QSS regime, the average sSFR is proportional to the specific accretion 
rate, sAR, with an absolute value that is insensitive to the SFR efficiency 
in a dynamical time $\epsilon$ and to the outflow mass-loading factor $\eta$.
The gas fraction is determined by $\epsilon$, and $\stv$ is driven by $\eta$.

\smallskip
At high redshifts, where the accretion is assumed to consist of gas only,
the simple toy model reproduces the observations rather well with no need
for fine tuning.
The specific SFR is predicted with no free parameters to be 
sSFR$\,\simeq [(1+z)/3]^{5/2} \Gyr^{-1}$, slightly higher than the cosmological
sAR, in general agreement with the rather noisy observed sSFR at $z=3-8$. 
The observed gas fraction constrains the SFR efficiency in a dynamical time 
to $\epsilon \simeq 0.02$.
The low $\stv$ indicated from observations requires an outflow mass-loading
factor of $\eta \simeq 1-3$, for a penetration efficiency of fresh 
gas into the galaxy of $\p = 0.5-1$ respectively.
Thus, the main features of galaxy evolution at high redshifts are
captured by the simplest toy model.

\smallskip
However, at $z \!\sim\! 2$, where stars are also accreted, through mergers,
there is a difficulty in matching the observations.
The model with the highest possible sSFR, where the outflows are fully 
recycled, falls short by a factor $\sim\! 3$ compared to the observed sSFR, 
and it overestimates $\stv$.
With $\eta \!\sim\! 3$, the model reproduces the latter but underestimates the
sSFR by an order of magnitude.
Thus, the toy model points at a robust challenge to theory at $z \!\sim\! 2$.

\smallskip
A potential way to ease the tension at $z\!\sim\!2$ is by massive recycling,
as in the case modeled by $\eta =0$ in the minimal bathtub model. 
A more sophisticated implementation of recycling in
the model may be required for a better match.
The missing population of quenched galaxies in the observed average sSFR
hints at an additional promising remedy. The galaxies where $\eta$ or $\fsa$
are higher than average drive the predicted average sSFR down, 
and the same galaxies are indeed likely to be quenched in reality.  
Once these are removed, the predicted sSFR would become higher by a factor of
2-3 and thus closer to the observed value.
Finally, the discrepancy at $z \!\sim\!2$ may be due to a breakdown of the 
QSS solution, e.g., by strong fluctuations in the accretion rate.
Alternatively, one could not deny the possibility of some fundamental flaw 
in the assumptions that lie at the basis of the bathtub toy model.
Identifying such a flaw would be very interesting.

\smallskip
One should conclude with a reminder that the bathtub toy model, at least in its
trivial form discussed here, is not meant to provide a detailed model that
perfectly matches the observations. It is useful for a simple and crude
qualitative study of the central elements of galaxy evolution, and for 
pointing out robust successes or major potential failures in reproducing 
certain observations, which should trigger a more detailed study of these 
issues. This has been demonstrated here by showing the robust match to 
observations at $z=3-8$, and by pointing out the non-trivial discrepancy 
at $z\!\sim\!2$.

\smallskip
On the other hand, the bathtub toy model can be extended to address
other major observables, such as the evolution of gas metallicity in
and outside galaxies and the scaling relations involving metallicity, 
as well as the hot gas content in the circum-galactic medium 
filling the halo and the associated with hot-mode accretion into the galaxy. 
Such extensions can be found, e.g., in 
\citet{kd12}, \citet{dave12}, and \citet{lilly13}, 
and they are shown to be good approximations
(Pipino et al.~2014, in preparation).

%%%%%%%%%%%%%
\section*{Acknowledgments}

This research has been supported by ISF grant 24/12,
by GIF grant G-1052-104.7/2009,
by a DIP grant,
by NSF grant AST-1010033,
and by the I-CORE Program of the PBC and
the ISF grant 1829/12.

%%%%%%%%%%%%%%%%%%%%%%%
%\appendix

%\section{Steady State: Approximate Solutions}
%\label{sec:app_ss}

%%%%%%%%%%%%%%%%%%%%%%%%%%%%%%%%%%%%%%%%%%%%%
\bibliographystyle{mn2e}
\bibliography{dekel_bathtub}
%%%%%%%%%%%%%%%%%%%%%%%%%%%%%%%%%%%%%%%%%%%%%

%\appendix
%\section{Useful Relations}
%\label{sec:useful}

\label{lastpage}
\end{document}